# Interplay between strong correlations and electronic topology in the underlying kagome lattice of $Na_{2/3}CoO_2$.

I.F. Gilmutdinov[1,4], R. Schönemann[2], D. Vignolles[1], C. Proust[1], I.R. Mukhamedshin[3,4], L. Balicas[2] and H. Alloul[3]

[1] Laboratoire National des Champs Magnétiques Intenses, 31400 Toulouse, France.

[2] National High Magnetic Field Lab, Florida State University, Tallahassee, Florida, 32310 USA

[3] Université Paris-Saclay, CNRS, Laboratoire de Physique des Solides, 91405, Orsay, France.

[4] Institute of Physics, Kazan Federal University, 420008 Kazan, Russia

## Abstract

Electronic topology in metallic kagome compounds is under intense scrutiny. We present transport experiments in $Na_{2/3}CoO_2$ in which the Na order differentiates a Co kagome sub-lattice in the triangular $CoO_2$ layers. Hall and magnetoresistance (MR) data under high fields give evidence for the coexistence of light and heavy carriers. At low temperatures, the dominant light carrier conductivity at zero field is suppressed by a $B$-linear MR suggesting Dirac like quasiparticles. Lifshitz transitions induced at large $B$ and $T$ unveil the lower mobility carriers. They display a negative $B^2$ MR due to scattering from magnetic moments likely pertaining to a flat band. We underline an analogy with heavy Fermion physics.

***Introduction:*** The influence of frustration of exchange on the magnetic properties of transition metal compounds has been investigated thoroughly in spin structures. The Herbertsmithite compound, whose $Cu^{2+}$ sites are ordered in a two-dimensional kagome structure, is considered as a good reference for a Quantum Spin Liquid, as no spin ordering has been detected in its ground state (1).

Attempts to synthesize doped metallic states in this kagome system by chemical substitutions have been so far unsuccessful (2). Those were motivated by a search for superconductivity in the phase diagram but also by an analogy with the honeycomb lattice such as that of graphene in which Dirac points are located (3). Local Density Approximation (LDA) calculations for a single orbital per site in weakly correlated kagome lattices also show the presence of Dirac or Weyl points as well as flat electronic bands (4). One wonders how such topological states would evolve in the presence of strong correlations. This led intensive searches for correlated kagome compounds e.g. in *3d* stannite materials like FeSn (5), CoSn (6) or $Co_3Sn_2S_2$ (7).

Here, we present an alternative approach based on the Na cobaltate compounds $Na_xCoO_2$ where the Co atoms are ordered on a triangular lattice. The originality of this system has been revealed by earlier NMR/NQR experiments. It was shown that the Na located between the $CoO_2$ layers (Fig. 1a) displays distinct orderings depending on the Na content (8) (9). The electrostatic incidence of the $Na^+$ ionic order induces a charge disproportionation of the Co sites (10). This has been evidenced in great detail in the case of the *x*=2/3 phase (11) (12) (13) in which a subset of Co sites remain in a $Co^{3+}$ state with filled non-magnetic $t_{2g}$ orbitals (Fig. 1b) while the complementary set of Co sites are ordered in a kagome sub-lattice having delocalized charge carriers (Fig. 1d). LDA calculations indicate that the Na order yields the minimum energy state for this compound (14), while LDA+*U* computations (15) demonstrate that a large coulomb interaction *U* is required to induce the disproportionation of the Co sites.

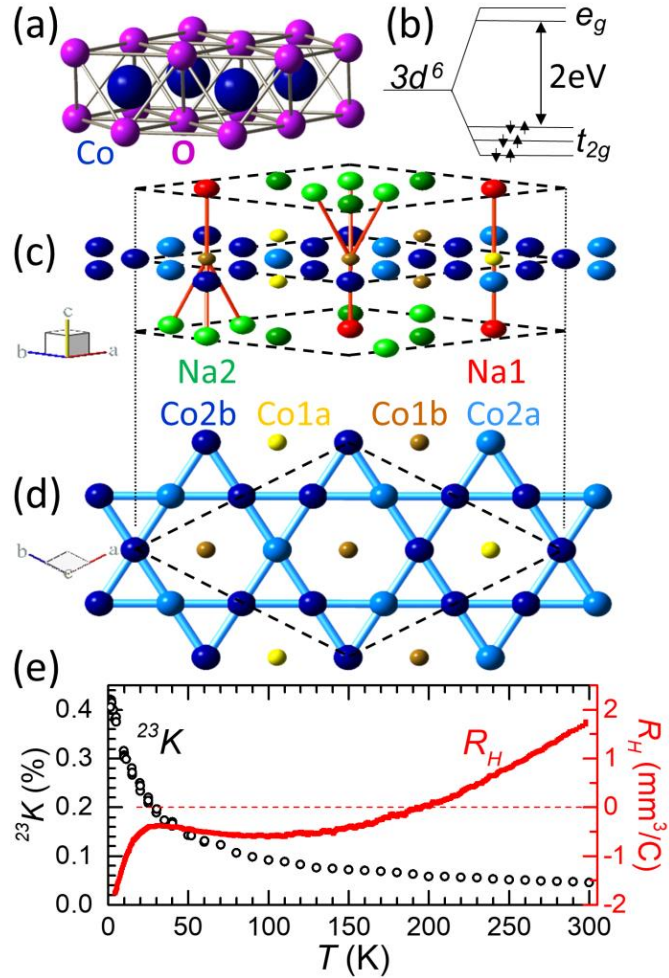

Fig. 1. (a) Structure of an isolated $CoO_2$ layer. (b) Splitting of the Co $3d$ levels induced by the crystal field in $Na_xCoO_2$ for $x=1$. (c) Differentiation for $x=2/3$ of the Co sites induced by the ordered stacking of Na above and below the $CoO_2$ layer. (d) 2D arrangement of the two types of sites in the Co plane (11). The $^{59}$Co NMR shift data gave evidence that Co1a and Co1b (yellow and brown) are non-magnetic $Co^{3+}$ sites with filled $t_{2g}$ levels as in (b). The $3d$ orbitals of the Co2a and Co2b sites (light and dark blue) arranged in a kagome sub-lattice are both nearly identically involved at the Fermi level(10) (16). (e) The $^{23}$Na NMR shift $^{23}K$ (left scale) monitors the spin susceptibility on those sites (associated with the hole doping of the correlated electronic bands of their kagome structure). It displays a Curie-Weiss like $T$ dependence (17) below 30 K concomitantly with the anomalous increase in magnitude of the Hall constant $R_H$ (right scale) (18).

Early experiments performed on samples with Na content near $x=2/3$ have indeed revealed singularly large values of the magnetic susceptibility (17) (19), specific heat(20) (21) and of $T^2$ dependence of the resistivity (22) which are obvious signs of strong electronic correlations. NMR shift measurements on the various Na and Co nuclear sites have given evidence that the local spin susceptibility $\chi$ of the Co2a and Co2b sites forming the kagome sub-lattice displays a large increase below 30 K (Fig. 1e), with respect to phases with a different Na content. Meanwhile, µSR experiments do not provide evidence for static magnetic order down to 0.1 K(23). Hence the local $\chi$ reaches a constant value only below $T$=1.5 K(17).

We have recently synthesized(24) high quality single crystals of the Na ordered $Na_{2/3}CoO_2$ phase with large residual resistivity ratios RRR=R(300 K)/R(1.5 K)~200. A change in the sign of the Hall-effect at 200 K was found to be followed below 30 K by a reproducible and unexpected large increase(18) in its negative magnitude (Fig. 1e). These modifications of the electronic properties indicate that the reconstruction of the Fermi Surface (FS) which occurs already above 200 K is followed by a large increase in electronic carrier mobility below 30 K.

Those results therefore underscore the need to perform detailed low $T$ band structure (BS) and FS studies on this specific kagome lattice material. As quantum oscillations (QOs) were discovered in uncontrolled quality samples(25) we have done high field measurements on our high-quality single crystals. We did not find QOs immediately in this phase but disclosed a series of new unexpected behaviors in the transport data.

We shall detail hereafter that an applied field of ~30 T in the ground state induces a change in the sign of the Hall effect, implying a major change in the electronic properties. We shall then underline the contrasting behavior of the magnetoresistance (MR) which changes sign and field dependence for increasing $T$. The negative $B^2$ dependence of the MR attributable to heavy carriers will be assigned to paramagnetic spin scattering in analogy to similar observations in the Heavy Fermion compounds. Comparisons with the multiband BS known for kagome compounds will lead us to suggest that the $B$ linear MR of the mobile carriers could be associated with Dirac/Weyl linear dispersing bands. In this peculiar metallic kagome compound the band structure topology therefore retains singular behaviors in the presence of strong correlations.

**Experimental results**

***Hall effect:*** Transport data was taken on a series of distinct samples and in two different high magnetic field facilities as detailed in the Supplemental Material (26). The small residual resistivity (~2 µΩ·cm) measured in our single crystals proves the low level of disorder for this $x$=2/3 phase. In preliminary studies done above 2 K the Hall resistivity $\rho_{xy}$ was found to be linear in $B$ below 9 T(18). But, when searching for QO at $T$=0.35 K under high magnetic fields at the Maglab in Tallahassee, we found that $\rho_{xy}$ goes through a minimum at ~15 T, and *becomes positive above 30 T* as displayed in Fig. 2a.

This drastic non-linear behavior only moderately changes when the temperature is increased up to $T$=5 K. Then $\rho_{xy}$ progressively becomes linear in $B$ and remains negative above 30 K within the experimental field range. Our extensive data set for the Hall constant $R_H$ (26) are summarized in the Fig. 2b. It provides evidence that $R_H$ is nearly $T$ and $B$ independent up to 1.5 K and 12 T but abruptly changes its sign above ~30 T and levels off at a positive value at the highest fields

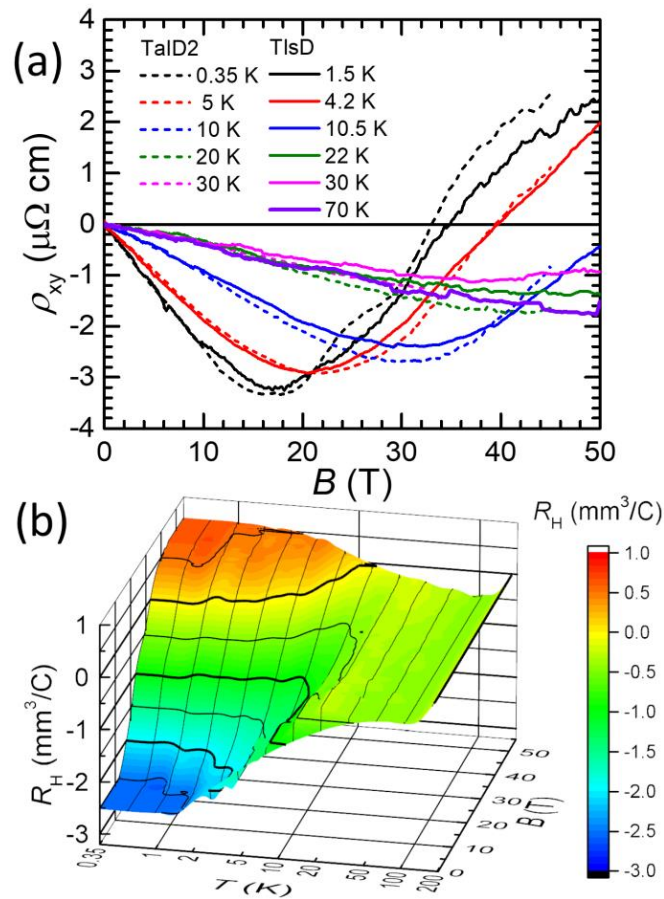

Fig. 2. (a) Hall resistivity $\rho_{xy}$ as a function of $B=\mu_0 H$ at various $T$ taken under DC applied fields $H$ for sample TalD2 and under pulsed fields for sample TlsD agrees perfectly. (b) The full dataset described in (26), section II are displayed as a 3D plot of the Hall constant $R_H$ versus $B$ and $T$ (log scale). $R_H$ becomes positive under large fields for $T$<10 K, while the low field behavior of Fig. 1e is seen to saturate below 1.5 K at its lower negative value (26).

This abrupt sign change of $R_H$ beyond 30 T implies a sharp reduction in electronic carrier density and/or mobility. So, assuming a Landé factor g=2, a Zeeman energy $g\mu_B B$~2 meV is sufficient to switch the transport from electron to hole-like carriers. On the contrary, the low field $R_H$ remains negative up to 200 K so that the electrons remain dominant well beyond 30 K. These results demonstrate that thermal and Zeeman energies have distinct incidences on the transport.

*Magnetoresistance:* In the first search for Shubnikov de Haas effect at T=0.35 K we did not find any indication for QOs at high frequencies which would be associated with large FS pockets. But as shown in Fig. 3a an unexpectedly large positive MR was detected (~100% at 10 T and 800 % at 45 T). At T=0.35 K it exhibits an initial *linear in field behavior*, while above 5 K it becomes *negative and follows a $B^2$ dependence*.

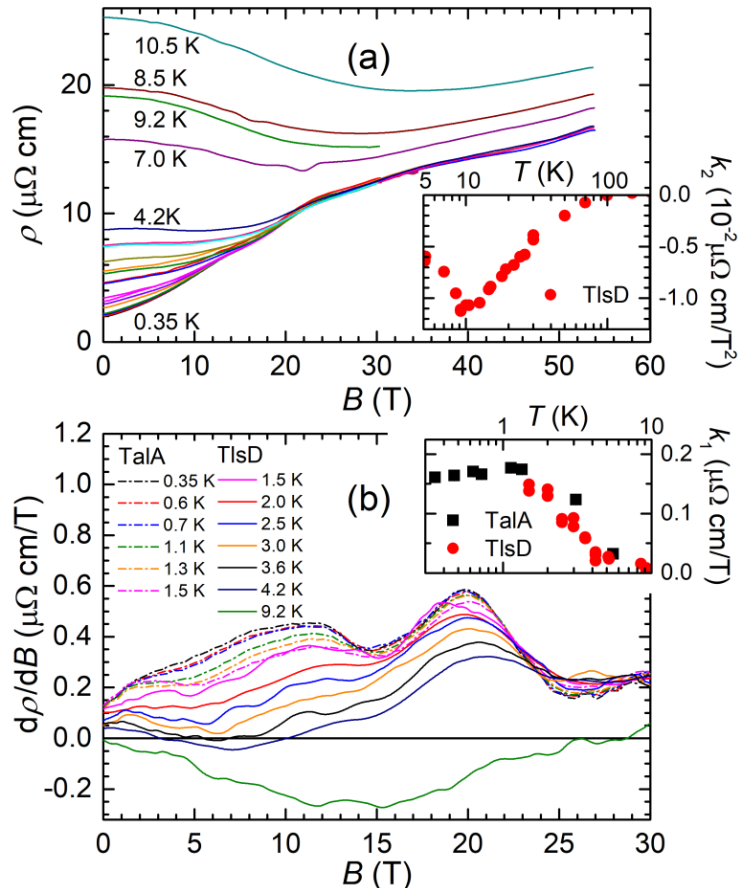

Fig.3 (a) Selected $\rho(B)$ curves taken from 0.35K to 1.5K for TalA sample and from 1.5K to 10.5K for TlsD. Details are available in (26), sec. II. (b) $d\rho(B)/dB$ displays maxima at 12T below 1.5K and at 20T below 5K. Insets: in (a) high T variation of the negative coefficient $k_2$ of the $B^2$ MR including data above 10K detailed in (26), sec. II. (b) Low T variation of the linear coefficient $k_1$ of the MR (see also (26), sec. V).

Other unusual behaviors also appear as the large $T$ variation of $\rho(B,T)$ displayed for $B=0$ T disappears up to 1.5 K for fixed field $B>12$ T and even up to 5 K for fixed $B>20$ T. Those two fields also appear as slight steps in $\rho(B)$ seen as maxima in the $d\rho(B)/dB$ curves of Fig. 3b. There the maximum at 20 T is seen for $T$ up to 5 K, while the one at 12 T disappears above 1.5 K. These derivative curves were found to be astonishingly reproducible on distinct sample batches (26).

The similar drastic loss of the ground state conductivity $\sigma=\sigma_{xx}$ which occurs beyond 20 T or beyond 5 K is illustrated in the 3D representation shown in Fig. 4. There $\sigma$ is shown in a logarithmic scale versus $B$ and $T$, with the latter also on a log scale. The corresponding linear representation of $\sigma$ is displayed in (26) where it is shown that for this kagome phase $\sigma \sim \rho^{-1}$. In Fig. 4, the absence of $T$ dependence of $\rho$ below 5 K for $B>20$ T is illustrated by the constant field curves in the yellow range.

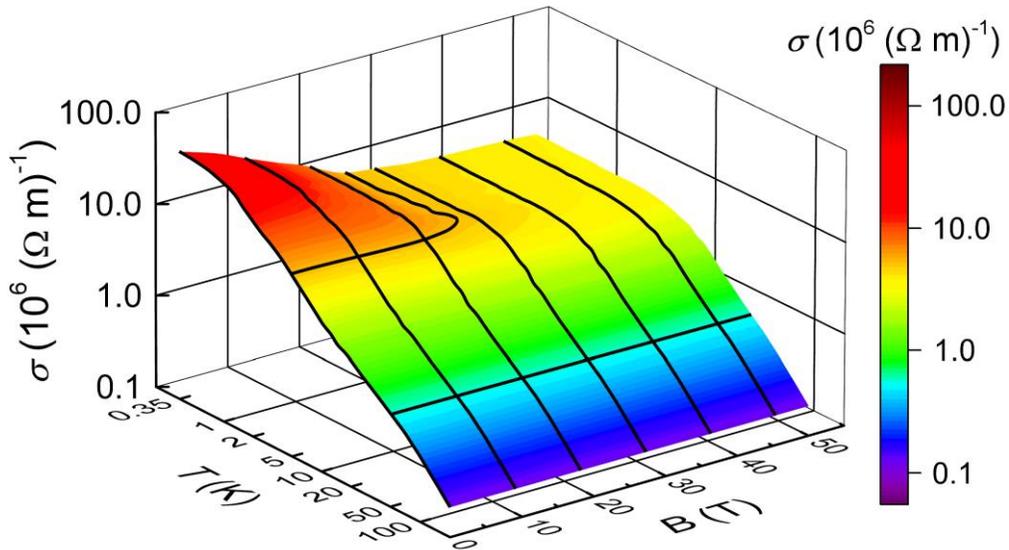

Fig. 4: 3D plot of the conductivity $\sigma(B,T)$ on a logarithmic scale highlighting the large increase which occurs below 5 K under small applied fields $B<20$ T. Here, data have been displayed up to T=100 K (log scale).

### Ground state multiband data analysis

The minimal model one might use to analyze the low $T$ data is a two-band model with hole and electron carriers at the Fermi level. For weakly interacting electrons in semiconductors or semi-metals(27) with quadratically dispersing bands the conductivities are $\sigma_h = n_h e \mu_h$ and $\sigma_e = n_e e \mu_e$ (where $n_h$, $n_e$,

$\mu_h$ and $\mu_e$ are the hole and electron carrier densities and mobilities). The classic equations for the conductivity and Hall constant, taking into account the Lorentz force acting on the carriers are(28):

$$\sigma = \frac{\sigma_h}{1+\mu_h^2 B^2} + \frac{\sigma_e}{1+\mu_e^2 B^2} \qquad (1)$$

$$R_H = \frac{1}{en_H} = \left[\frac{\sigma_h \mu_h}{1+\mu_h^2 B^2} - \frac{\sigma_e \mu_e}{1+\mu_e^2 B^2}\right]\sigma^{-2} \qquad (2)$$

Measurements of $\sigma$, $R_H$ and of the positive $B^2$ MR would allow us to extract the $T$ dependencies of the carrier densities and mobilities assuming a charge neutrality condition. This approach has been found to hold even in some correlated metals such as the Fe pnictides(29).

In the present compound the NMR data implies a total number of 0.44 holes per Co atomic site of the kagome lattice, so eqs. (1) and (2) can be complemented by $n_h-n_e$=0.44. But this model does not explain the linear MR, which points towards singular band properties. We notice however in Fig. 4 that in the low $T$ range the applied field $B$ suppresses the conductivity of the high mobility electronic carriers and discloses that of the hole carriers. Therefore in the $T$=0 B~0 ground state for which the Hall effect remains constant $\sigma$ splits into two contributions $\sigma_e$ = 40.10$^6$($\Omega$ m)$^{-1}$ and $\sigma_h$=(1/4)$\sigma_e$. We can use there eqs. (1) and (2) with $R_H$=-2.5 mm$^3$/C that is $n_H$=0.26 per Co site of the underlying kagome lattice, which yields $n_e$~0.16, $n_h$=0.60 and $\mu_e/\mu_h$=15.

The main conclusion of this simple two band analysis in terms of "effective"(30) carrier concentrations and mobilities is that *the holes are heavy carriers*. But this model cannot explain the $B$-linear contribution to the MR and has many other shortcomings. For instance, the steps in $\rho(B)$ at 12 T and 20 T suggest the occurrence of successive Lifshitz transitions. One could consider that such transitions eliminate above 20 T all electron pockets from the FS. But in Fig. 2 the variation of $R_H$ with increasing $B$ is smooth beyond 12 T and only changes sign above 30 T. This implies that at least an electron pocket remains above 20 T so that the multiband aspect persists in the high field range.

**Magneto-transport properties as functions of *T***

The loss of conductivity seen up to 5 K in Fig 4 is concomitant with the large decrease of $|R_H|$ and could be due to a mobility decrease for some of the electronic pockets. One should distinguish the initial increase of $\rho(B=0)$ thermally driven by ground state excitations from its evolution beyond 5 K.

***Fermi liquid properties:*** As can be seen in Fig. 4, the zero-field conductivity decreases rapidly with increasing $T$. We show in section IV of (26) that $\rho(T)=\rho_0+A\,T^2$ up to T=1.5 K, with similar $A$ values for samples with slightly different $\rho_0$ values. Such a variation had been seen as well on a less characterized sample in which the $x=2/3$ ordered phase was apparently dominant(22). In a single band approach this had been interpreted as Fermi liquid behavior with a $T^2$ electron-electron scattering rate. Here, although many bands contribute to $\sigma$, we may conclude that the Fermi liquid scenario applies to the highly mobile electrons which dominate the transport at low temperatures. But beyond 20 T, under a fixed field $\sigma$ becomes $T$ independent up to 5 K while heavy carriers dominate the conductivity. Thus, these carriers display weaker intra-band scattering than the high mobility electrons. As we found (26) that the $T=0$ resistivity for $B>20$ T is sample independent, their mobility might then be governed by an intrinsic process such as a quantum inter-band scattering (31).

***Magnetoresistance:*** We can clearly see in Fig. 3 the linear in $B$ contribution to the MR which appears in the $B=0$ limit of the isothermal $d\rho/dB$. Fits of the data up to 5 T with a second order polynomial dependence $\rho(B,T)=\rho_0(T)+k_1 B+k_2 B^2$ shows that $k_1$ decreases with increasing $T$ and vanishes for $T\sim 5$ K as seen in the inset of Fig. 3b. This linear contribution is therefore intrinsically related to the conductivity of the high mobility electronic carriers. On the contrary beyond 5 K the MR is dominated by the negative $k_2 B^2$ contribution and $d\rho/dB$ becomes $B$-linear in Fig. 3b. This MR detected upon suppression of the mobile electron conductivity is therefore associated with the heavy carriers. The reproducibility of $k_1$ and $k_2$ data displayed in section IV of (26) for distinct samples allows us to underline that those MR coefficients are intrinsic to the clean metallic phase.

### Discussion

We have disclosed here a large contrast in the transport properties among the various carrier bands in $Na_{2/3}CoO_2$. The local Curie Weiss like susceptibility (Fig. 1e) has previously established the existence of large electronic correlations in the underlying Co kagome band structure. A hallmark of this behavior is the negative quadratic MR due to spin scattering, usually absent in weakly correlated materials.

The reproducibility of the NMR and transport data (26) allows us to exclude *a priori* disorder or extrinsic impurities for this local moment behavior. This negative MR is mostly effective on the low mobility carriers and is therefore reminiscent of the situation encountered in some Heavy Fermion Kondo lattice compounds. There the loss of coherence of the 4$f$ Kondo electrons influence the transport governed by the 5$d$ bands (32). Here, the interplay stands between bands built from the multiplet of Co 3$d$ sublevels. A similar proposal has been done for Hund differentiated Fe pnictides. (33)

This analogy in the incidence of the magnetic scattering contrasts however with the difference in most other magneto-transport properties. In the tetragonal 122 Fe pnictide the multiband analysis could be performed and matched with ARPES experiments (29), while quite weak magnetic contributions to the MR were detected (34). The $Na_{2/3}CoO_2$ metal surprisingly does not superconduct at low $T$ and displays a series of original transport behaviors which have been so far individually observed in specific materials. The linear MR is usually associated with linear dispersing bands, partial filling of the lowest Landau level of small FS pockets or kinks in the Fermi surface (35). The detected steps of the conductivity imply the occurrence of low energy features near the Fermi level so that Lifshitz like transitions are observed upon moderate increases of $B$ or $T$. The insensitivity to intrinsic disorder also raises fundamental questions about the respective incidences of intra and interband scattering.

An interpretation of these phenomena requires a specific investigation of the actual band structure of $Na_{2/3}CoO_2$ in which the kagome structure is certainly of major importance. We have indicated in (18) that ARPES experiments will be difficult to perform on the surface of Na ordered samples. As theoretical input is also still lacking, we may use as guidance the LDA calculations done for a single orbital per kagome lattice site. Those yield two conical Dirac bands and a flat band which cross at the Brillouin zone center (4). Such bands have been disclosed by ARPES experiments in the metallic kagome materials known so far, such as the Co stannites (5)(6)(7) or the recently discovered $AV_3Sb_5$ (A=alkali) (36). Therefore, it appears logical to anticipate that a similar situation occurs in the present case. With 0.44 hole per kagome site the Fermi level should occur within the flat band and the transport should depend markedly on the refined band structure modifications at the band crossing points.

As the conductivity of the large $\mu_e$ carriers is suppressed with a linear MR, one may speculate that those are associated with partially filled Dirac or Weyl linearly dispersing bands, for which a linear MR is expected (37).

The two steps suppression of these carriers with increasing $B$ or $T$ could indicate a slight energy splitting of the two conical bands. They would be depleted in turn by Lifshitz transitions due to chemical potential shifts that would disclose the low mobility carriers. The latter would then be associated with partly filled flat bands that are expected to display behaviors akin to localized states (38). One could alternatively wonder whether the mobile carriers could depart from their Fermi liquid behavior above the apparent Lifshitz transitions and become then incoherent and more localized.

To conclude we provided evidence indicating that topology, frustration, and correlated electron physics are entangled in this well characterized kagome material. Such an interplay has only been proposed to occur in few compounds. It is not blurred here by concomitant magnetic or structural phase transitions.

We hope that our results will stimulate realistic theoretical efforts including the role of Hund's and Spin Orbit couplings. Some of the open questions might as well be answered experimentally in future high field runs and through comparisons with the other kagome compounds.

## Acknowledgements

We thank S. Benhabib for her help in experimental runs in LNCMI and acknowledge exchanges with A. Subedi, L. de Medici, M. Civelli, M-O. Goerbig, K. Behnia, B. Fauqué and V. Brouet. The crystal growth, XRD, low field magnetic and transport measurements were carried out at the Federal Center of Shared Facilities of Kazan Federal University. Travel of HA and IG between Orsay, Kazan and Toulouse has been financially supported by an "Investissement d'Avenir" allowance from the Labex PALM (No. ANR-10-LABX-0039-PALM). Collaborative work between HA and LB have been supported by an ICAM travel grant between Orsay and Tallahassee funded by the Gordon and Betty Moore foundation QuantEmX program.
L.B. is supported by the US-DOE, BES program through award DE-SC0002613. The National High Magnetic Field Laboratory is supported by the National Science Foundation through NSF/DMR-1644779 and the State of Florida. We also acknowledge support by the LNCMI-CNRS, member of the European Magnetic Field Laboratory (EMFL).

# Supplemental Material for

# Interplay between strong correlations and electronic topology in the underlying kagome lattice of Na$_{2/3}$CoO$_2$.

I.F. Gilmutdinov[1,4], R. Schönemann[2], D. Vignolles[1], C. Proust[1],
I.R. Mukhamedshin[3,4], L. Balicas[2] and H. Alloul[3]

[1] Laboratoire National des Champs Magnétiques Intenses, 31400 Toulouse, France.

[2] National High Magnetic Field Lab, Florida State University, Tallahassee, Florida, 32310 USA

[3] Université Paris-Saclay, CNRS, Laboratoire de Physique des Solides, 91405, Orsay, France.

[4] Institute of Physics, Kazan Federal University, 420008 Kazan, Russia

# I - *Methods and Samples*

In order to investigate the electronic structure modifications induced by field and temperature in the *x*=2/3 cobaltate phase, we performed measurements in two high field facilities with complementary experimental ranges. We have taken data at the Maglab in Tallahassee from 0.35 K up to 40 K in He$^3$ cryostats in the continuous magnetic field facility (DC Field) in different resistive magnets as well as in the hybrid 45 T magnet. The 1.5 K to 200 K temperature range was explored in a He$^4$ cryostat in the Toulouse LNCMI pulse magnetic field facility up to 60 T. Data have been taken in Kazan on sample KznD in a dilution refrigerator down to 0.1 K and in a PPMS up to 300 K in fields below 9 T. The resistivity measurements were taken with AC current techniques with four or six contacts on the samples. The AC frequency was below 50 Hz in the experiments performed in the Maglab and about 51.6 kHz in the experiments performed in LNCMI.

Different samples batches were prepared along the lines detailed in references (18) and (24) in the main text. The main characteristics of the samples used in the present experiments are listed in table I.

In order to measure transport properties, samples with the following approximate dimensions were prepared: 3.0 x 1.5 x 0.1 mm for measurements in DC magnetic field, and 2.0 x 0.5 x 0.05 mm for experiments in pulsed magnetic field. We used silver wires (25 μm diameter) and silver paint DuPont 4929N to contact the samples. This allowed us to make contacts having a low resistance, smaller than 2 Ohms.

The contacts configuration for $R_{xx}$ and $R_{xy}$ measurements performed in DC and pulsed magnetic fields is shown below (Fig. S0). Transport properties of the sample KznD were measured using the van der Pauw method. In all experiments the current was applied along the *ab*-plane of the crystals and the magnetic field *B* along the *c*-axis. The current and voltage contacts were covering the sample surface on both sides, and also the side of the sample in order to short the *c*-axis. $R_{xx}$ and $R_{xy}$ measurements were performed in positive and negative magnetic fields $B_{UP}$ and $B_{DW}$ respectively. In order to remove the contribution of the spurious voltage due to contact misalignment, $R_{xx}$ and $R_{xy}$ were calculated according to the following equations:

$$R_{xx} = \frac{R(B_{UP})+R(B_{DW})}{2} \qquad R_{xy} = \frac{R(B_{UP})-R(B_{DW})}{2}$$

**Table I:** List of the samples used with label names starting by a reference to the field facility: Tal (DC field, Maglab, Tallahassee), Tls (pulsed field, LNCMI, Toulouse) and KznD (Dilution fridge with 9 T magnet, Kazan). For a given sample in the following columns + symbols indicate the quantity measured that is $R_{xx}$, $R_{xy}$, then the maximum magnetic field which depends on the used facility. In the last columns we display the resistivity values measured at 0.35 K or 1.5 K and the resistivity ratio $RRR = \rho\,(300\,K)\,/\,\rho\,(1.5\,K)$.

| Sample name | $R_{xx}$ | $R_{xy}$ | Max $B$ (T) | $\rho$ (µΩ cm) $T = 0.35$ K | $\rho$ (µΩ cm) $T = 1.5$ K | RRR |
|---|---|---|---|---|---|---|
| TalA | + | - | 34.5 | 2 | 3.2 | - |
| TalB1 | + | - | 33 | 1.8 | 3 | - |
| TalB3 | + | - | 33 | - | - | - |
| TalB4 | + | - | 33 | 2.3 | 3.5 | - |
| TalB5 | + | + | 33 | 2.4 | 3.6 | - |
| TalD2 | + | + | 45 | 2.8 | - | - |
| TlsB | + | - | 65.5 | 2.4 | 3.5 | - |
| TlsD | + | + | 55 | - | 3.4 | 221 |
| KznD | + | + | 9 | 1.8 | 3.1 | 276 |

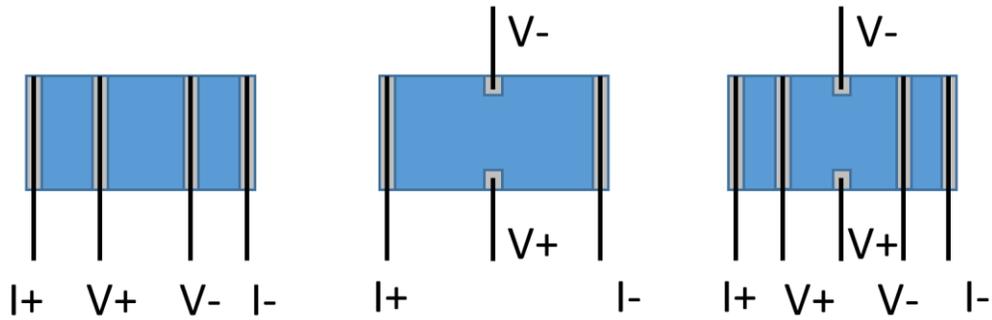

Fig. S0. Schematic drawing of the contacts configuration used for $R_{xx}$ and $R_{xy}$ transport measurements. The c-axis of the crystals is orthogonal to the plane of the drawing. I+ and I- are current leads, V+ and V- are contacts used to measure the voltages.

## II - $R_H$ and MR supplementary data

Evidence was given in the main text for the reproducibility of $\rho_{xy}$ data taken in two distinct high field facilities on two different samples (Fig. 2a). We detail here some of the larger set of data which allowed us to construct the 3D plot of Fig. 2b.

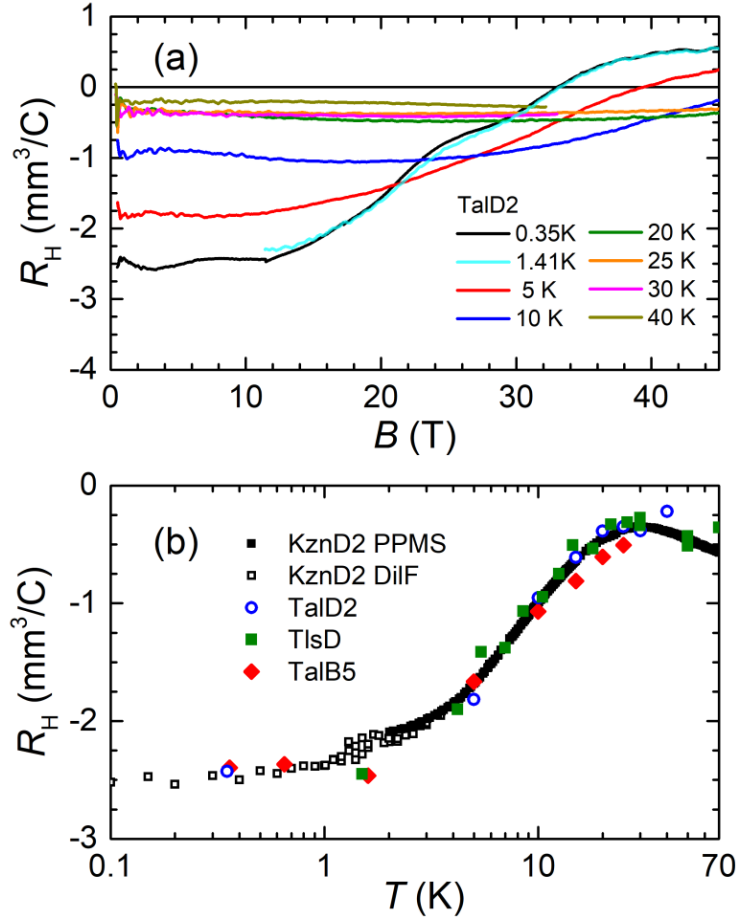

Fig. S1. **(a)** $R_H(B)$ as a function of field as deduced from the $\rho_{xy}$ data of the main text for sample TalD2. Data curves between 0.35 K and 1.41 K are superimposed and have been omitted **(b)** $T$ variation of the low field data for $R_H$ obtained in Kazan, Maglab and Toulouse. The measured $R_H$ values for distinct samples have been scaled to compensate the lack of accuracy on their geometrical factors. The log scale for $T$ allows to emphasize that $R_H(B)$ reaches a nearly constant value within experimental accuracy below 1.5 K for the KznD and TalB5 samples.

We present in Fig. S1a the $R_H$ values for the sample TalD2 deduced from $\rho_{xy}$ data including that of Fig. 2a of the main text. One can see there that $R_H$ is field independent within experimental accuracy below about 10 T but changes sign at low $T$ beyond 30 T. In this plot we only kept the two extreme data curves between 0.35 K and 1.41 K out of the five ones obtained at intermediate *T* in the DC hybrid magnet at the Maglab as all curves accurately overlap. This explains the low *T* plateau and cliff like behavior displayed in the 3D plot.

In Fig. S1b one can see that the *T* dependences of the low field data for $R_H$ taken in DC field and in pulsed fields on distinct sample batches are found to be in perfect agreement, though the experimental noise is larger as might be expected for the pulsed field experiments (as seen as well in Fig. 2a). The lack of *T* dependence below 1.5 K within experimental accuracy is enlightened by the data taken for *B*=9 T in the dilution fridge in Kazan on sample KznD. These results with complementary intermediate temperature data have been used to construct the 3D Fig. 2b of the main text.

Let us complete also the information on the magneto-conductivity which can be deduced from the MR data. Here we present in Fig. S2 the data for 1/ρ vs *B*. It complements the resistivity data of Fig. 3a of the main text as it allows to report the higher *T* data. One can see there that reliable data from 4.2 K to 10 K in the pulse field facility were somewhat difficult to obtain within the limited allotted experimental time. The small difference displayed between the data at *T* quoted as 8.5 K and 9.2 K in Fig. 3(a) of the main text is much less apparent here. Those data were taken on different days of the experiment. The fact that they are not exactly on the same regular trend illustrates the actual experimental accuracy of the temperature determination in the pulse field experiment. This is specific to this *T* range obtained in this set up by cooling with a He4 exchange gas. The field variation is seen however to be identical for the two data curves.

While in zero field the conductivity is σ=1/ρ, in an applied field *B* the resistivity and conductivity tensors become non-diagonal due to the transverse Hall effect resistivity $\rho_{xy}$. So, the diagonal conductivity is given by

$$\sigma = \sigma_{xx} = \frac{\rho_{yy}}{\rho_{xx}\rho_{yy} - \rho_{xy}\rho_{yx}} \tag{s1}$$

Our samples are single crystals in the *c* direction but the Na order is a mosaic in the *ab* plane so we only measure an average transverse conductivity and are insensitive to any eventual in plane anisotropy so that

$$\sigma = \sigma_{xx} = \frac{\rho}{\rho^2 - \rho_{xy}^2} \tag{s2}$$

Here we could measure accurately both $\rho$ and $\rho_{xy}$ in the same facility and on the very same sample only in the case of few samples (table I). The tensor analysis could be strictly done only for those limited results. But the reproducibility of the data found in these experiments allowed us to perform the analysis using interpolated results for $\rho$ such as those shown in 3D for $\rho_{xy}$ in Fig. 2b. With the interpolated data for $\rho_{xy}$ from Fig. 3a and Fig. S2 we obtained the 3D plot for $\sigma$ reported in the main text (Fig. 4).

Let us point out that in the present case $\rho_{xy}^2 < \rho^2$ for all applied fields and temperatures and the maximum relative correction with respect to $\sigma_0 = 1/\rho$ is at most of ~25% for B~15 T at the lowest $T$.

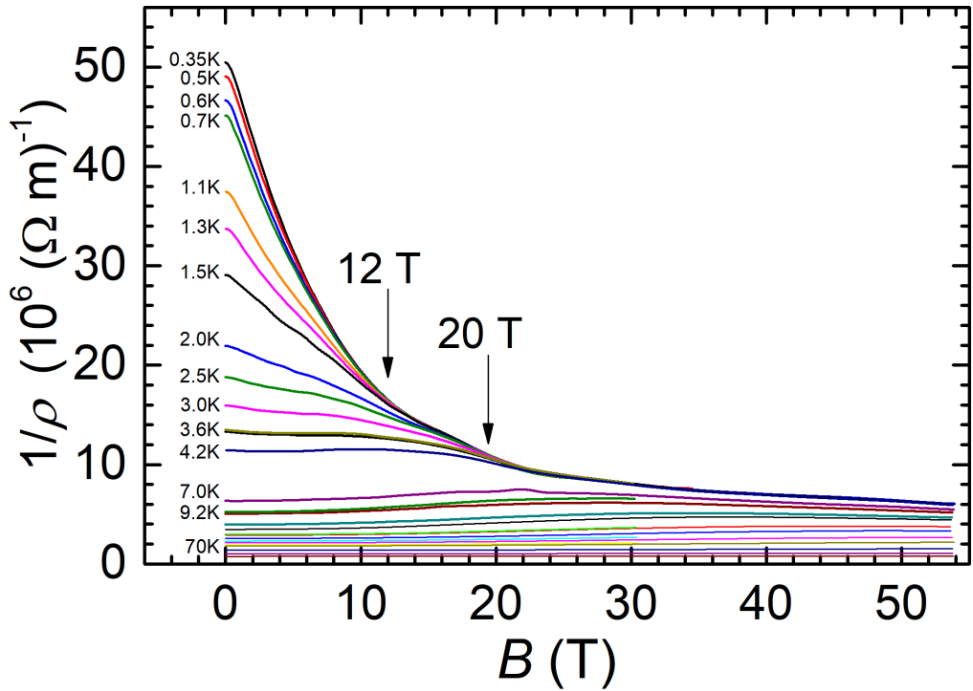

Fig. S2. Linear plot of $1/\rho$ (B) at various temperatures. Here one can perceive the large increase of the low $T$ conductivity due to the high mobility carriers as well as the singular field points where the curves merge at 12 T and 20 T.

## III - *Reproducibility of the MR singularities*

We have shown in the main text that the MR displays singularities which are better seen in the d$\rho$/dB curves. We found that these singularities are present for all samples studied. The reproducibility of the data can be seen in Fig. S3a, where we have plotted the d$\rho$/dB data as a function of field taken on samples of distinct batches at the base temperature of 0.35 K at the Maglab. In this plot the raw data was re-scaled by numerical factors not far from unity due to the lack of accuracy on the determinations of the geometrical factors (sizes and contact positions on the single crystal samples). Data were taken at both 0.35 K and 1.5 K for some samples and the same scaling factors were found to apply to both temperatures, which indicates that the $T$ dependence of $\rho$ is also reproducible as will be confirmed in the next sections.

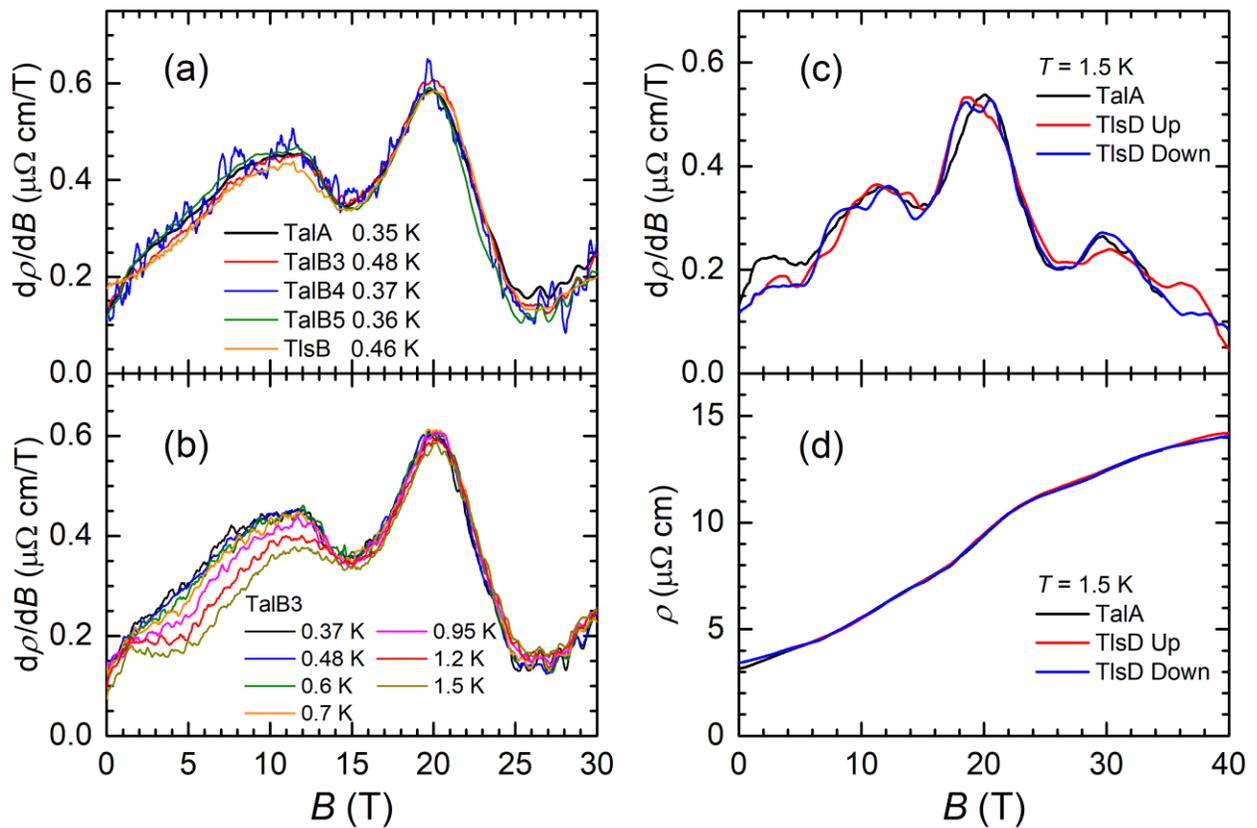

Fig. S3. (a) Comparison between the derivatives of the $\rho(B)$ curves taken at $T$~0.35 K for various samples. (b) Accurate low $T$ variation of d$\rho(B)$/dB. (c) Scaling at 1.5 K of the d$\rho(B)$/dB data taken for positive and negative pulsed field on the TlsD sample compared to the one obtained under DC fields on the TalA sample. (d) Corresponding $\rho(B)$ for these samples at T=1.5 K.

We also show in Fig. S3b the $d\rho/dB$ data taken at the Maglab on a distinct sample for a series of temperatures below 1.5 K. One can see in this figure that the low $T$ dependence of the two singularities is quasi identical to that displayed in Fig. 3 of the main text for which a larger $T$ range was investigated. But in this study at the Maglab an excellent S/N ratio was achieved so one can see that the 20 T singularity remains absolutely unchanged up to 1.5 K while the anomaly at 12 T decreases by almost 30% at 1.5 K. The latter apparently also shifts to higher fields with increasing $T$ as also seen on Fig. 3b. These singularities are therefore differently affected by increasing $T$ which indicates that they cannot be interpreted as QOs. Instead they ought to be electronic Lifshitz phase transitions which might progressively merge with increasing $T$ up to 5 K.

In order to explore a larger range of temperatures and fields we have been led to compare the data taken under DC fields at the Maglab and under pulsed fields at the LNCMI-Toulouse. Though we have compared data taken in a large $T$ range, systematic comparisons done at 1.5 K on distinct samples are illustrated in Fig. S3. As can be seen in Fig. S3c the data for the TalA and TlsD samples can be scaled perfectly, although as expected the experimental noise is somewhat larger for the data collected in the pulsed field facility. Using this scaling allows to demonstrate the very good reproducibility of the $\rho(B)$ data taken at 1.5 K as shown on Fig. S3d. This allowed us to combine the data taken for these two samples which are presented in Fig. 3 within the main text.

# IV – *T dependence of the resistivity*

In this section we compare the low $T$ behavior of the resistivity measured on various samples. We attempted fits to the power law $\rho=\rho_0+AT^n$ for $0<T<1.5$ K. For the sample TlsB for which accurate data was taken at 70 Hz lock-in frequencies in zero field we obtained a fit with the exponent $n=1.92(1)$ close to a $T^2$ law. We therefore fixed $n=2$ for the data analysis of all our samples.

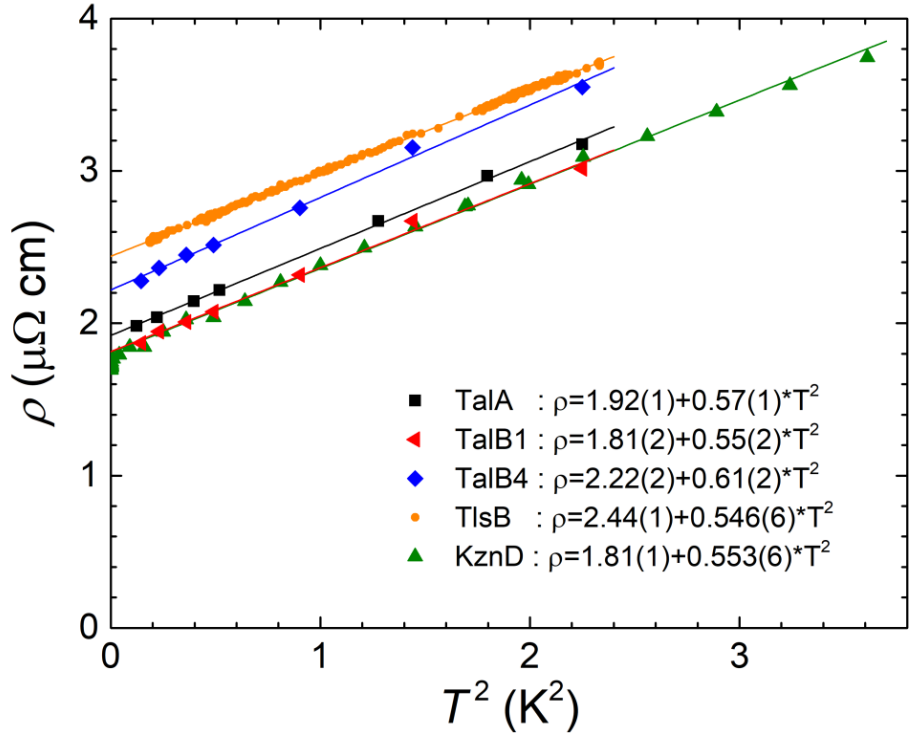

Fig. S4. Experimental $T$ dependence of the $B=0$ resistivity for five samples with results of the statistical fits with a $T^2$ variation.

As displayed in Fig. S4, the accurate fits allowed us to demonstrate the reproducibility of the coefficient $A$. Note that the residual resistivity only slightly depends on the sample batch or the experimental set-up. This $T^2$ variation might be interpreted as a Fermi liquid behavior, assuming a single band with a fixed number of carriers. In that case, a $T^2$ electron-electron scattering rate adds to a $T$ independent impurity scattering rate. In the main text we show that the transport properties are associated with a multiband behavior. The dominant contributions of light electron bands to $\sigma$ at low field are apparently eliminated in two steps at 12 T and 20 T as underlined in the main text and in section III. Their contribution dominates the $T$ variation of the conductivity below 5 K while $\sigma$(20 T), the contribution of the heavier carrier bands is nearly $T$ independent. $\sigma_1 = \sigma(0T) - \sigma(20T)$ is dominated by the overall contribution of the lighter electron bands. We have consistently checked that the $T^2$ Fermi liquid behavior applies for $\rho_1 = 1/\sigma_1$ with a coefficient $A$ only slightly larger than that obtained for the global resistivity $\rho$.

One is tempted to consider that these electronic bands should also be the origin of the dominant contributions to the macroscopic low $T$ specific heat and spin susceptibility measured in this cobaltate phase. The $A$ value is nearly as large as values obtained from heavy Fermions as in Ref. (22) in the main text, where a slightly larger value $A = 1$ $\mu\Omega$ cm K$^{-2}$ is quoted for a sample which was not well characterized. The actual specific heat measured in cobaltate samples with a similar Na content would indicate a very large value of the Kadowaki-Woods empirical ratio $A/\gamma^2$ as given in Ref. (22) in the main text.

But as emphasized in the main text the observed initial $B$ linear MR implies that the electronic carriers cannot be associated with a standard parabolic band. A better analysis of these data will need to await experimental or theoretical determinations of the band structure in this underlying kagome phase.

## V - *B dependences of the MR*

Let us consider now in some detail the reproducibility of the $B$ dependence of the MR which is described in the main text.

*Low temperatures and light carriers*:

We report here the $T$ dependence of $k_1$, the $B$ linear coefficient of the MR. The results for $k_1(T)$ as obtained through second order polynomial fits to our data for various samples are reported in Fig. S5. It can be seen there that the fits are reproducible among the various samples and are typically independent of $T$ up to 1.5 K, then $k_1$ decreases continuously with increasing $T$, and vanishes at about 5 K where the negative $B^2$ MR becomes dominant.

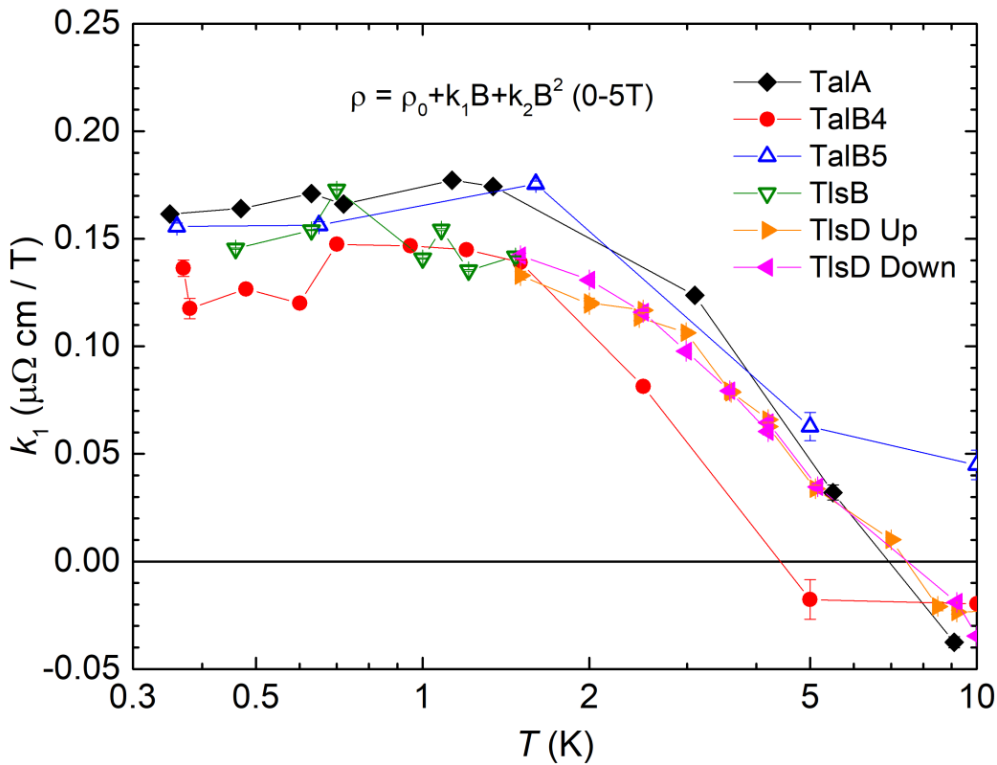

Fig. S5. Experimental $T$ dependence of the linear term in the $B$ dependence of the MR plotted versus $T$ on a log scale.

*Heavy bands at higher T:*

We compare in Fig. S6 the MR data taken at T ~ 10 K for a series of samples. One finds there that the data follows a nearly identical negative $B^2$ dependence. The data taken at different magnet facilities agree quite well within an experimental accuracy of at most 20%. This reproducibility of both the $k_2$ slope and the zero field resistivity values leads us to conclude that the magnetic response is not associated with extrinsic disorder such as defects.

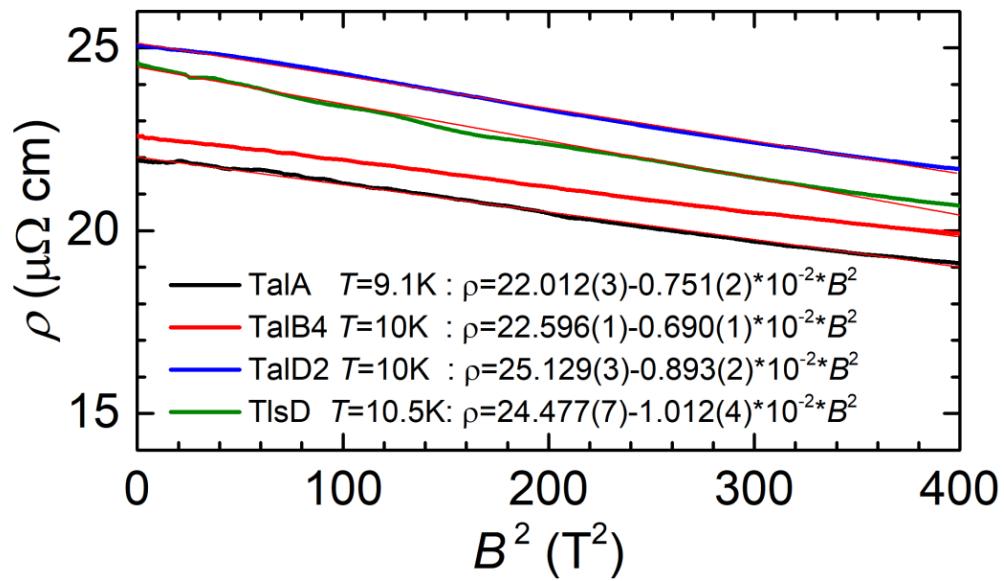

Fig. S6. Resistivity al T~10 K as a function of the field squared for four distinct samples with results of the statistical fits to a $B^2$ dependence.